\documentclass[11pt,superscriptaddress,aps,prd,preprint]{revtex4}
\usepackage{amsmath}
\usepackage{graphicx}

\makeatletter
\usepackage[T1]{fontenc}
\usepackage{amsmath}
\usepackage{amssymb}
\usepackage{graphicx}

\usepackage{slashed}

\newcommand{\bea}{\begin{eqnarray}}
\newcommand{\eea}{\end{eqnarray}}

\begin{document}

\title{Casimir effect and Stefan-Boltzmann law at finite temperature in a Friedmann-Robertson-Walker universe}

\author{A. F. Santos}\email[]{alesandroferreira@fisica.ufmt.br}
\affiliation{Instituto de F\'{\i}sica, Universidade Federal de Mato Grosso,\\
78060-900, Cuiab\'{a}, Mato Grosso, Brazil}

\author{S. C. Ulhoa}\email[]{sc.ulhoa@gmail.com}
\affiliation{International Center of Physics, Instituto de F\'isica, Universidade de Bras\'ilia, 70910-900, Bras\'ilia, DF, Brazil} \affiliation{Canadian Quantum Research Center,\\ 
204-3002 32 Ave Vernon, BC V1T 2L7  Canada}

\author{Faqir C. Khanna\footnote{Professor Emeritus - Physics Department, Theoretical Physics Institute, University of Alberta\\
Edmonton, Alberta, Canada}}\email[]{khannaf@uvic.ca}
\affiliation{Department of Physics and Astronomy, University of Victoria,\\
3800 Finnerty Road Victoria, BC, Canada}

\begin{abstract}

A spatially flat Friedmann-Robertson-Walker background with a general scale factor is considered. In this space-time, the energy-momentum tensor of the scalar field with a general curvature coupling parameter is obtained. Using the Thermo Field Dynamics (TFD) formalism the Stefan-Boltzmann law and the Casimir effect at finite temperature are calculated. The Casimir effect at zero temperature is also considered. The expansion of the universe changes these effects. A discussion of these modifications is presented.

\end{abstract}

\maketitle

\section{Introduction}

The present observational data leads to a universe that is in a period of accelerated expansion and appears to be spatially flat \cite{Riess, Per, Spergel, Teg}. Quantum effects in an expanding universe are an interesting topic that should be investigated in a cosmological context \cite{rev1}. More precisely thermal effects and the Casimir effect are of particular interest. Quantum effects, related to temperature, have their origin in Stefan-Boltzmann law for black-body radiation.  There are several examples in our universe as stars that behave like black-bodies.  On a cosmological scale, the cosmic background radiation has a temperature associated with it whose explanation may be linked to quantum effects on the origin of the universe. If an expanding universe is considered, then new contributions to quantum effects are expected.

The Casimir effect is a direct manifestation of the physical reality due to the quantum vacuum. The original work was proposed by Casimir in 1948 \cite{Casimir}. Numerous theoretical and experimental studies have been carried out \cite{Review, Milton}, based on this work. The Casimir effect plays an important role in several fields of physics such as quantum field theory, gravitation and cosmology, condensed matter physics, atomic and molecular physics, among others. Initially, the Casimir effect was predicted for the electromagnetic field. However, now it is calculated for any quantum field. A scalar field may yield the vacuum.

In this paper, the Casimir effect and Stefan-Boltzmann law for a scalar quantum field with a general curvature coupling parameter is calculated. A spatially flat Friedman-Robertson-Walker (FRW) space-time is assumed as a geometrical background. There are several studies about the Casimir effect due to a scalar field coupled to gravity in different cosmological backgrounds. For example, the topological Casimir effect in compactified cosmic string space-time has been considered \cite{Bez1}, scalar Casimir effect in a high-dimensional cosmic space-time has been analyzed \cite{Bez2}, scalar Casimir effect in a linearly expanding universe has been investigated \cite{Bez3}, scalar Casimir densities and forces for parallel plates in cosmic string space-time have been calculated \cite{Bez4}, Casimir effect for parallel plates in a Friedmann-Robertson-Walker universe has been developed \cite{Bez5}, among others. Here the Casimir effect at finite temperature and the Stefan-Boltzmann law for the scalar field coupled to gravity in the FRW space-time are obtained. The temperature effects are introduced using the Thermo Field Dynamics (TFD) formalism.

The temperature effects in a thermal quantum field theory may be introduced using the imaginary or real time formalism. The imaginary time formalism or Matsubara formalism \cite{Matsubara} is based on a substitution of time, $t$, by a complex time, $i\tau$. The real time formalism consists of two approaches, the closed time path formalism \cite{Schwinger} and the Thermo Field Dynamics (TFD) formalism \cite{Umezawa1, Umezawa2, Umezawa22, Khanna1, Khanna2}. In both formalisms, it is possible to preserve the dependence of time together with the effects of temperature. Here the TFD formalism is used. This procedure depends on the doubling of the Hilbert space, composed of the original and a tilde space (dual space), using the Bogoliubov transformations.  This doubling of Hilbert space is defined by the tilde conjugation rules, associating each operator, say $a$, in ${\cal S}$ to two operators in ${\cal S}_T$, with ${\cal S}_T={\cal S}\otimes \tilde{\cal S}$, where $\tilde{\cal S}$ is the tilde space. The temperature effects are introduced by the Bogoliubov transformation that implies a rotation involving the two spaces.

This paper is organized as follows. In section II, a brief introduction to TFD is presented. In section III, the theoretical model is discussed. The energy-momentum tensor for the scalar field coupled to gravity is obtained. The line element that describes the FRW universe, i.e. a flat and expanding universe, is given. In sections IV, some applications are developed. The Stefan-Boltzmann law is obtained at finite temperature. The Casimir effect at zero and finite temperature is obtained. In section V, some concluding remarks are presented.

\section{Thermo Field Dynamics (TFD) formalism}

In this section a brief introduction to TFD formalism is presented. In this formalism a thermal vacuum, $|0(\beta) \rangle$ ($\beta=\frac{1}{k_BT}$, with $T$ being the temperature and $k_B$ the Boltzmann constant) is constructed. From this thermal vacuum, the main idea is to interpret the statistical average of an arbitrary operator $A$, as the expectation value in a thermal vacuum, i.e., $\langle A \rangle=\langle 0(\beta)| A|0(\beta) \rangle$. For this interpretation two conditions are imposed: (i) the Hilbert space ${\cal S}$ is doubled, i.e., the thermal space is defined as ${\cal S}_T={\cal S}\otimes \tilde{\cal S}$, where $\tilde{\cal S}$ is the dual (tilde) Hilbert space, and (ii) the Bogoliubov transformation is used. The Bogoliubov transformation introduces thermal effects through a rotation between tilde ($\tilde{\cal S}$) and non-tilde (${\cal S}$) operators. 

By taking arbitrary operators ${\cal D}$ and $\tilde{\cal D}$ in Hilbert space ${\cal S}$ and tilde space $\tilde{\cal S}$ respectively, the Bogoliubov transformation is
\bea
\left( \begin{array}{cc} {\cal D}(k, \alpha)  \\\eta \tilde {\cal D}^\dagger(k,\alpha) \end{array} \right)={\cal B}(\alpha)\left( \begin{array}{cc} {\cal D}(k)  \\ \eta\tilde {\cal D}^\dagger(k) \end{array} \right),
\eea
where $k$ is the 4-momentum and $\eta = -1$ for bosons and $\eta = +1$ for fermions. The Bogoliubov transformation, ${\cal B}(\alpha)$, is defined as
\bea
{\cal B}(\alpha)=\left( \begin{array}{cc} u(\alpha) & -v(\alpha) \\
\eta v(\alpha) & u(\alpha) \end{array} \right),
\eea
with $u^2(\alpha)+\eta v^2(\alpha)=1$. The functions $u(\alpha)$ and $v(\alpha)$, are related to the Bose distribution, and are given as
\bea
v^2(\alpha)=(e^{\alpha\omega_k}-1)^{-1}, \quad\quad u^2(\alpha)=1+v^2(\alpha),\label{phdef}
\eea
with $\omega_k=k_0.$
The $\alpha$ parameter is assumed as the compactification parameter defined by $\alpha=(\alpha_0,\alpha_1,\cdots\alpha_{D-1})$, with $D$ being the space-time dimension. The effect of temperature is described by the choice $\alpha_0\equiv\beta$ and $\alpha_1,\cdots\alpha_{D-1}=0$. Then the functions $u^2(\beta)$ and $v^2(\beta)$ are fermion or boson distribution functions. In general, $\alpha$ may be associated with any physical quantities. 

In this paper a field theory on the topology $\Gamma_D^d=(\mathbb{S}^1)^d\times \mathbb{R}^{D-d}$ with $1\leq d \leq D$ is considered. Here $d$ is the number of compactified dimensions. This establishes a formalism such that any set of dimensions of the manifold $\mathbb{R}^{D}$ are compactified, where the circumference of the $nth$ $\mathbb{S}^1$ is specified by $\alpha_n$. Here three different topologies are used. (i) The topology $\Gamma_4^1=\mathbb{S}^1\times\mathbb{R}^{3}$, where $\alpha=(\beta,0,0,0)$. (ii) The topology $\Gamma_4^1$ with $\alpha=(0,0,0,i2b)$. (iii) The topology $\Gamma_4^2=\mathbb{S}^1\times\mathbb{S}^1\times\mathbb{R}^{2}$ with $\alpha=(\beta,0,0,i2b)$. Here $2b$ corresponds to the length of the circumference $\mathbb{S}^1$.

The TFD formalism is used to introduce $\alpha$-parameter which is the compactification parameter for the propagator of the theory. Let's take the scalar field as an example. Then the propagator is written as
\bea
G_0^{(AB)}(x-x';\alpha)&=&i\langle 0(\alpha)| \tau[\phi^A(x)\phi^B(x')]| 0(\alpha)\rangle,\nonumber\\
&=&i\int \frac{d^4k}{(2\pi)^4}e^{-ik(x-x')}G_0^{(AB)}(k;\alpha),
\eea
where $A, B=1, 2$ represent the doubled space. Then
\bea
G_0^{(AB)}(k;\alpha)={\cal B}^{-1}(\alpha)G_0^{(AB)}(k){\cal B}(\alpha),
\eea
with
\bea
G_0^{(AB)}(k)=\left( \begin{array}{cc} G_0(k) & 0 \\
0 & \eta G^*_0(k) \end{array} \right),
\eea
and
\bea
G_0(k)=\frac{1}{k^2-m^2+i\epsilon},
\eea
begin the usual scalar field propagator with the mass $m$. $G^*_0(k)$ is the conjugate complex of $G_0(k)$.

Then the Green function is
\bea
G_0^{(11)}(k;\alpha)=G_0(k)+\eta\, v^2(k;\alpha)[G^*_0(k)-G_0(k)],
\eea
where $v^2(k;\alpha)$ is the generalized Bogoliubov transformation \cite{GBT} which is given as
\bea
v^2(k;\alpha)=\sum_{s=1}^d\sum_{\lbrace\sigma_s\rbrace}2^{s-1}\sum_{l_{\sigma_1},...,l_{\sigma_s}=1}^\infty(-\eta)^{s+\sum_{r=1}^sl_{\sigma_r}}\,\exp\left[{-\sum_{j=1}^s\alpha_{\sigma_j} l_{\sigma_j} k^{\sigma_j}}\right],\label{BT}
\eea
with $d$ being the number of compactified dimensions, $\eta=1(-1)$ for fermions (bosons), $\lbrace\sigma_s\rbrace$ denotes the set of all combinations with $s$ elements.

\section{The model - Scalar field coupled to gravity}

The main objective is to obtain the energy-momentum tensor for the scalar field non-minimally coupled to the gravitational background. The action that describes a scalar field $\phi(x)$ non-minimally coupled to gravity is
\bea
S=\frac{1}{2}\int d^4x\sqrt{-g}\left(g^{\mu\nu}\partial_\mu\phi(x)\partial_\nu\phi(x)-m^2\phi(x)^2-\xi R\phi(x)^2\right),
\eea
where $g$ is the metric determinant and $\xi$ is the coupling parameter to the curvature scalar $R$.

By varying the action with respect to the scalar field, the equation of motion is obtained as
\bea
\left(\partial_\mu\partial^\mu+m^2+\xi R\right)\phi(x)=0.
\eea
It should be noted that the limit $\xi=0$ represents the decoupling between the scalar and gravity field. This is a mandatory constraint in order to split both fields even for the particular solution $\dot{a}=0$.

In order to calculate the Stefan-Boltzmann law and the Casimir effect using the TFD formalism, the energy-momentum tensor for this model is obtained. It is defined as
\bea
T_{\gamma\rho}=-\frac{2}{\sqrt{-g}}\frac{\delta S}{\delta g^{\gamma\rho}}.
\eea
Using this
\bea
\frac{\delta \sqrt{-g}}{\delta g^{\gamma\rho}}&=&-\frac{1}{2}\sqrt{-g}g_{\gamma\rho},\\
\frac{\delta g^{\mu\nu}}{\delta g^{\gamma\rho}}\partial_\mu\phi(x)\partial_\nu\phi(x)&=&\partial_\gamma\phi(x)\partial_\rho\phi(x),\\
\frac{\delta( \xi R\phi(x)^2)}{\delta g^{\gamma\rho}}&=&\xi\left(R_{\gamma\rho}+g_{\gamma\rho}\Box-\partial_\gamma\partial_\rho\right)\phi(x)^2,
\eea
the energy-momentum tensor is written as
\bea
T_{\gamma\rho}&=&\frac{1}{2}g_{\gamma\rho}\partial^\mu\phi(x)\partial_\mu\phi(x)-\frac{1}{2}g_{\gamma\rho}m^2\phi(x)^2-\partial_\gamma\phi(x)\partial_\beta\phi(x)\nonumber\\&+&\xi\left(R_{\gamma\rho}-\frac{1}{2}g_{\gamma\rho}R+g_{\gamma\rho}\Box-\partial_\gamma\partial_\rho\right)\phi(x)^2.
\eea

To avoid divergences, the energy-momentum tensor is written at different space-time points, i.e.,
\bea
T_{\gamma\rho}(x)&=&\lim_{x'\rightarrow x}\tau\Biggl\{\frac{1}{2}g_{\gamma\rho}\partial^\mu\phi(x)\partial_\mu\phi(x')-\frac{1}{2}g_{\gamma\rho}m^2\phi(x)\phi(x')-\partial_\gamma\phi(x)\partial_\beta\phi(x')\nonumber\\&+&\xi\left(R_{\gamma\rho}-\frac{1}{2}g_{\gamma\rho}R+g_{\gamma\rho}\Box-\partial_\gamma\partial_\rho\right)\phi(x)\phi(x')\Biggl\},\label{EMT}
\eea
where $\tau$ is the time ordering operator. Using the commutation relation
\bea
[\phi(x),\partial'^\rho\phi(x')]=in_0^\rho\delta({\vec{x}-\vec{x'}}),
\eea
where $n_0^\rho=(1,0,0,0)$ is a time-like vector and $\partial^\rho\theta(x_0-x_0')=n_0^\rho\,\delta(x_0-x_0')$. The energy-momentum tensor becomes
\bea
T_{\gamma\rho}(x)&=&\lim_{x'\rightarrow x}\Bigl\{\Gamma_{\gamma\rho}\tau\left[\phi(x)\phi(x')\right]-I_{\gamma\rho}\delta(x-x')\Bigl\},
\eea
with
\bea
\Gamma_{\gamma\rho}=\frac{1}{2}g_{\gamma\rho}\partial^\mu\partial_\mu-\partial_\gamma\partial_\beta+\xi\left(R_{\gamma\rho}-\frac{1}{2}g_{\gamma\rho}R+g_{\gamma\rho}\Box-\partial_\gamma\partial_\rho\right)
\eea
and
\bea
I_{\gamma\rho}=-\frac{i}{2}g_{\gamma\rho}n_0^\mu\,n_{0\mu}+in_{0\gamma}n_{0\rho}.
\eea

In order to obtain important physical quantities, the vacuum expectation value of the energy-momentum tensor is 
\bea
\left\langle T_{\gamma\rho}(x)\right\rangle=\lim_{x'\rightarrow x}\Bigl\{i\Gamma_{\gamma\rho}G_0(x-x')-I_{\gamma\rho}\delta(x-x')\Bigl\},
\eea
with $G_0(x-x')$ being the scalar field propagator defined as
\bea
iG_0(x-x')=\left\langle 0\left|\tau[\phi(x)\phi(x')]\right| 0 \right\rangle.
\eea

To use the TFD formalism, the duplicate notation for the energy-momentum tensor is
\bea
\left\langle T_{\gamma\rho}^{(AB)}(x;\alpha)\right\rangle=\lim_{x'\rightarrow x}\Bigl\{i\Gamma_{\gamma\rho}G_0^{(AB)}(x-x')-I_{\gamma\rho}\delta(x-x')\delta^{(AB)}\Bigl\},
\eea
To obtain a physical (renormalized) energy-momentum tensor, the usual Casimir prescription is,
\bea
{\cal T}_{\gamma\rho}(x;\alpha)=\left\langle T_{\gamma\rho}^{(AB)}(x;\alpha)\right\rangle-\left\langle T_{\gamma\rho}^{(AB)}(x)\right\rangle.
\eea
Then
\bea
{\cal T}_{\gamma\rho}(x;\alpha)=\lim_{x'\rightarrow x}\Bigl\{i\Gamma_{\gamma\rho}\overline{G}_0^{(AB)}(x-x';\alpha)\Bigl\},
\eea
where
\bea
\overline{G}_0^{(AB)}(x-x';\alpha)=G_0^{(AB)}(x-x';\alpha)-G_0^{(AB)}(x-x').\label{Green}
\eea

Some applications of ${\cal T}_{\gamma\rho}(x;\alpha)$ for a cosmological background are considered. A background geometry in a spatially flat 4-dimensional FRW space-time is used in an expanding universe. The line element is described by
\bea
ds^2=dt^2-a^2(t)\left(dx^2+dy^2+dz^2\right),
\eea
where $a(t)$ is the scale factor. For this space-time, the non-zero components of the Ricci tensor are
\bea
R_{00}&=&-3\frac{\ddot{a}}{a},\\
R_{ij}&=&\delta_{ij}\left(a\ddot{a}+2\dot{a}^2\right),
\eea
and the Ricci scalar is
\bea
R=-6\left[\frac{\ddot{a}}{a}+\left(\frac{\dot{a}}{a}\right)^2\right],
\eea
where dot means derivative with respect to time. In addition, the Einstein tensor is defined as
\bea
G_{\gamma\rho}=R_{\gamma\rho}-\frac{1}{2}g_{\gamma\rho}R
\eea
and its non-zero components are
\bea
G_{00}=3\left(\frac{\dot{a}}{a}\right)^2\quad\quad\quad \mathrm{and} \quad\quad\quad G_{ii}=-\left(2a\ddot{a}+\dot{a}^2\right)
\eea
with $i=1, 2, 3$. The Einstein equation is coupled to the scalar field by means of the coupling constant $\xi$, in order to find a classical solution one should note that the scale factor $a(t)$ depends on the scalar field which is constrained by the choice of $\xi$. Thus the scale factor itself is not independent on the coupling constant. These quantities will be used to calculate the Casimir effect and the Stefan-Boltzmann Law for scalar field
at finite temperature in an expanding FRW universe.

\section{Thermal and Casimir effects}

Here the Stefan-Boltzmann law and the Casimir effect at zero and finite temperature are calculated in a flat FRW universe, that serves as a geometric background.

\subsection{Stefan-Boltzmann law at finite temperature}

To obtain effect of finite temperature, it is defined as $\alpha_0\equiv\beta$ and $\alpha_1,\cdots, \alpha_{D-1}=0$. Then the generalized Bogoliubov transformation is
\bea
v^2(\beta)=\sum_{l_0=1}^{\infty}e^{-\beta k^0l_0}.\label{BT1}
\eea
The Green function is 
\bea
\overline{G}_0(x-x';\beta)=2\sum_{l_0=1}^{\infty}G_0(x-x'-i\beta l_0n_0),\label{GF1}
\eea
where $n_0=(1,0,0,0)$ and $\overline{G}_0(x-x';\beta)\equiv \overline{G}_0^{(11)}(x-x';\beta)$, that is the physical component. Then the energy-momentum tensor is
\bea
{\cal T}^{(11)}_{\gamma\rho}(x;\beta)&=&2i\lim_{x'\rightarrow x}\Bigl\{\Gamma_{\gamma\rho}\sum_{l_0=1}^{\infty}G_0(x-x'-i\beta l_0n_0)\Bigl\}.
\eea

By taking $\gamma=\rho=0$ we get
\bea
{\cal T}^{(11)}_{00}(x;\beta)&=&2i\lim_{x'\rightarrow x}\Bigl\{\Gamma_{00}\sum_{l_0=1}^{\infty}G_0(x-x'-i\beta l_0n_0)\Bigl\},\label{11}
\eea
where
\bea
\Gamma_{00}&=&-\frac{1}{2}\partial_0\partial'_0-\frac{1}{2a^2(t)}\left(\partial_1\partial'_1+\partial_2\partial'_2+\partial_3\partial'_3\right)+\xi\left[G_{00}-\frac{1}{a^2(t)}\left(\partial_1\partial'_1+\partial_2\partial'_2+\partial_3\partial'_3\right)\right].\label{00}
\eea

\begin{figure}[htbp]
	\centering
		\includegraphics[width=0.60\textwidth]{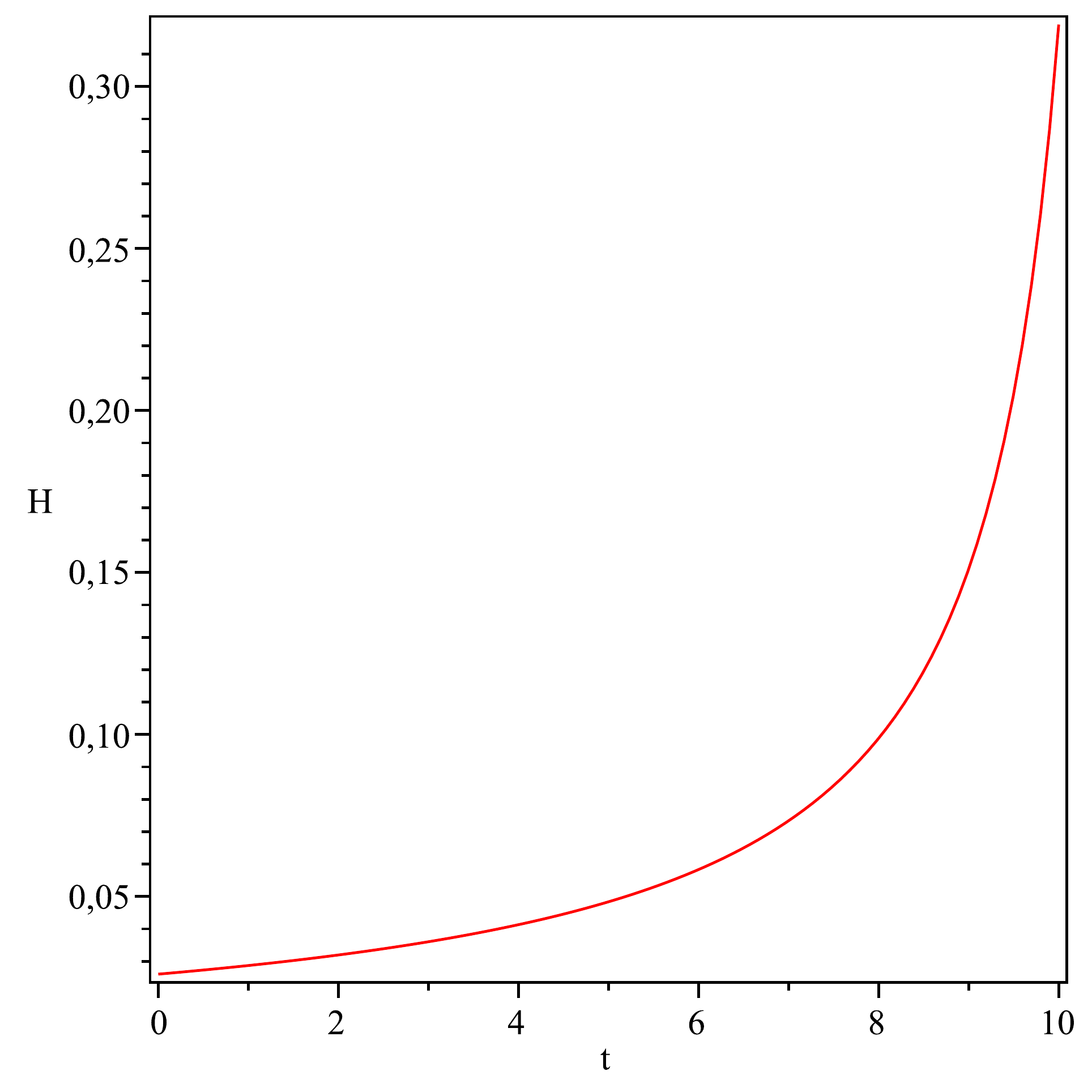}
		\caption{ The graphic of $H= \frac{\dot{a}}{a}$ for $\xi=0.1$ is plotted. This solution is obtained when the initial conditions, $\phi(0)=5$ and $\dot\phi(0)=0.1$, for the scalar field are used. In addition this solution is obtained for a massless scalar field in a flat, $k=0$, Universe.}
	\label{fig1}
\end{figure}

In the flat FRW space-time the scalar field propagator is
\bea
G_0(x-x')=-\frac{i}{(2\pi)^2}\frac{1}{(x-x')^2},\label{scp}
\eea
with
\bea
(x-x')^2&=&g_{\mu\nu}(x-x')^\mu(x-x')^\nu\nonumber\\
&=&(t-t')^2-a^2(t)(x-x')^2-a^2(t)(y-y')^2-a^2(t)(z-z')^2.
\eea
In the TFD formalism, the Bogoliubov transformation is used, then the scalar field propagator, Eq. (\ref{scp}), becomes
\bea
G_0(x-x'-i\beta l_0n_0)=-\frac{i}{(2\pi)^2}\frac{1}{(x-x'-i\beta l_0n_0)^2},\label{GF}
\eea
where
\bea
(x-x'-i\beta l_0n_0)^2&=&(t-t'-i\beta l_0)^2-a^2(t)(x-x')^2-a^2(t)(y-y')^2-a^2(t)(z-z')^2.
\eea
Using equations (\ref{00}) and (\ref{GF}), the energy-momentum tensor, eq. (\ref{11}), becomes
\bea
{\cal T}^{(11)}_{00}(T)=\frac{\pi^2}{30}T^4+\xi\left[\frac{\pi^2}{30}T^4-\frac{1}{4}\left(\frac{\dot{a}}{a}\right)^2T^2\right].
\eea
This is the Stefan-Boltzmann law for the scalar field in an expanding FRW universe at finite temperature. It is important to note that: (i) the first term is the usual Stefan-Boltzmann law for a massless scalar field; (ii) the second term is a surface term; and (iii) the third term is due to the expansion of the universe. In addition, the expansion of the universe may decrease the energy density as presented in FIG. \ref{fig1}. As a matter of fact the energy can be negative for a given universe`s expansion, that suggests non-dissipation of phenomena at the beginning of the Universe.  This is because the attempt to dissipate a negative energy makes it more and more negative.  Thus, the energy of a system with negative energy does not tend to zero due to its loss. It is important to note that this scenario occurs when the time is close to zero. Furthermore, it is a unique result of TFD, since in the latter term there is a combined effect of temporal evolution and temperature.

\subsection{Casimir effect at zero temperature}

For Casimir effect, the topology $\Gamma_4^1=\mathbb{S}^1\times\mathbb{R}^{3}$, with $\alpha=(0,0,0,i2b)$ is considered. This leads to the compactification along the $z$-coordinate. Then  the Bogoliubov transformation is
\bea
v^2(b)=\sum_{l_3=1}^{\infty}e^{-i2b k^3l_3}\label{BT2}.
\eea
In this case the Green function becomes
\bea
\overline{G}_0(x-x';b)=2\sum_{l_3=1}^{\infty}G_0(x-x'-2b l_3n_3)\label{GF2}
\eea
with $n_3=(0,0,0,1)$. Taking $\gamma=\rho=0$ the energy-momentum tensor is given as
\bea
{\cal T}^{(11)}_{00}(x;b)&=&2i\lim_{x'\rightarrow x}\Bigl\{\Gamma_{00}\sum_{l_3=1}^{\infty}G_0(x-x'-2b l_3z)\Bigl\}.\label{111}
\eea
Using eq. (\ref{00}), the Casimir energy for scalar field in an expanding FRW universe is
\bea
{\cal T}^{(11)}_{00}(b)=-\frac{\pi^2}{1440b^4a^4}+\xi\left[-\frac{\pi^2}{1440b^4a^4}-\frac{\dot{a}^2}{16b^2a^4}\right].
\eea
It is important to note that the standard Casimir energy for massless scalar field (i.e., the case with $\xi=0$) is modified by the scale factor. Then the geometric background changes the Casimir energy.

\begin{figure}[htbp]
	\centering
		\includegraphics[width=0.60\textwidth]{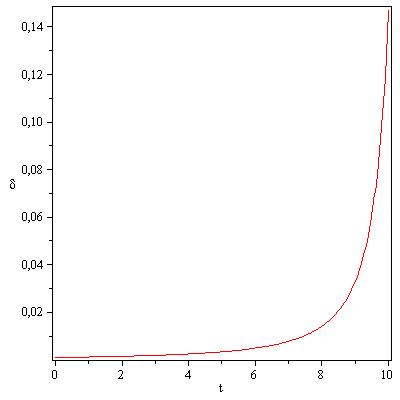}
		\caption{The graphic of $\delta= \left(\frac{\dot{a}}{a}\right)^2+ 2\left(\frac{\ddot{a}}{a}\right)$ for $\xi=0.1$ is plotted. This solution is obtained when the initial conditions, $\phi(0)=5$ and $\dot\phi(0)=0.1$, for the scalar field are used. In addition this solution is obtained for a massless scalar field in a flat, $k=0$, Universe.}
	\label{fig2}
\end{figure}

The Casimir pressure is calculated similarly by taking $\gamma=\rho=3$. Then the energy-momentum tensor becomes
\bea
{\cal T}^{(11)}_{33}(x;b)&=&2i\lim_{x'\rightarrow x}\Bigl\{\Gamma_{33}\sum_{l_3=1}^{\infty}G_0(x-x'-2b l_3z)\Bigl\},\label{33}
\eea
with
\bea
\Gamma_{33}&=&-\frac{a^2(t)}{2}\partial_0\partial'_0+\frac{1}{2}\left(\partial_1\partial'_1+\partial_2\partial'_2-\partial_3\partial'_3\right)+\xi\left[G_{33}-a^2(t)\partial_0\partial'_0+\partial_1\partial'_1+\partial_2\partial'_2\right].\label{G33}
\eea
Then the Casimir pressure is
\bea
{\cal T}^{(11)}_{33}(b)=-\frac{\pi^2}{480b^4a^2}+\xi\left[-\frac{\pi^2}{480b^4a^2}+\frac{\dot{a}^2+2a\ddot{a}}{48b^2a^2}\right].
\eea
It is important to note that, due to the expansion of the universe, there may be a transition to a positive value of the Casimir pressure. A repulsive Casimir effect is a possibility in an expanding universe as presented in FIG. \ref{fig2}. Such a repulsive effect may even be responsible for an accelerated expansion since there is an explicit time dependence.

\subsection{Casimir effect at finite temperature}

For Casimir effect at finite temperature, consider the topology $\Gamma_4^2=\mathbb{S}^1\times\mathbb{S}^1\times\mathbb{R}^{2}$ with $\alpha=(\beta,0,0,i2b)$. Then two compactifications, the time and the other along the $z$-coordinate are explored. Then the generalized Bogoliubov transformation becomes
\bea
v^2(\beta,b)=\sum_{l_0=1}^\infty e^{-\beta k^0l_0}+\sum_{l_3=1}^\infty e^{-i2bk^3l_3}+2\sum_{l_0,l_3=1}^\infty e^{-\beta k^0l_0-i2bk^3l_3},\label{BT3}
\eea
where the first two terms are associated with the Stefan-Boltzmann law and the Casimir effect at zero temperature and the third term provides the combined effect of temperature and spatial compactification. Then the Green function is
\bea
\overline{G}_0(x-x';\beta,b)&=&4\sum_{l_0,l_3=1}^\infty G_0\left(x-x'-i\beta l_0n_0-2bl_3n_3\right).\label{GF3}
\eea

For $\gamma=\rho=0$ the Casimir energy at finite temperature in an expanding FRW universe is 
\bea
{\cal T}^{(11)}_{00}(\beta,b)&=&-\frac{2a^2}{\pi^2}\sum_{l_0,l_3=1}^\infty\frac{[(2bl_3)^2a^2-3(\beta l_0)^2]}{a^2[(2bl_3)^2a^2+(\beta l_0)^2]^3}\nonumber\\
&+&\xi\Biggl\{-\frac{2a^2}{\pi^2}\sum_{l_0,l_3=1}^\infty\frac{[(2bl_3)^2a^2-3(\beta l_0)^2]}{a^2[(2bl_3)^2a^2+(\beta l_0)^2]^3}-\frac{3\dot{a}^2}{\pi^2a^2[(2bl_3)^2a^2+(\beta l_0)^2]}\Biggl\}
\eea
and the Casimir pressure at finite temperature ($\gamma=\rho=3$) is
\bea
{\cal T}^{(11)}_{33}(\beta,b)&=&-\frac{2a^2}{\pi^2}\sum_{l_0,l_3=1}^\infty\frac{[3(2bl_3)^2a^2-(\beta l_0)^2]}{a^2[(2bl_3)^2a^2+(\beta l_0)^2]^3}\nonumber\\
&+&\xi\Biggl\{-\frac{2a^2}{\pi^2}\sum_{l_0,l_3=1}^\infty\frac{[3(2bl_3)^2a^2-(\beta l_0)^2]}{a^2[(2bl_3)^2a^2+(\beta l_0)^2]^3}-\frac{\dot{a}^2+a\ddot{a}}{\pi^2[(2bl_3)^2a^2+(\beta l_0)^2]}\Biggl\}.
\eea
As in previous cases, the Casimir energy and pressure at finite temperature are modified due to the expansion of the universe.

\section{Conclusions}

The scalar field coupled to gravity in the spatially flat FRW background is considered. In order to introduce temperature effects the TFD formalism is used. A field theory on the topology $\Gamma_D^d=(\mathbb{S}^1)^d\times \mathbb{R}^{D-d}$ is studied. This establishes a formalism such that any set of dimensions of the manifold $\mathbb{R}^{D}$ can be compactified. Inclusion of the TFD formalism leads to effects due to temperature. When the time dimension is compactified, the effect of temperature emerges. In this case, the Stefan-Boltzmann law at finite temperature in a FRW space-time is calculated. The results show that the expansion of the universe changes the energy density. The Casimir effect at zero temperature is obtained when the z-component is compactified. This result depends on the expansion of the universe. Due to the expansion of the universe, a transition for a positive value of the Casimir force, i.e., a repulsive Casimir effect, is a possibility in this model. In addition, the Casimir effect at finite temperature is calculated. This result emerges from the compactification of time and one spatial dimension. The main results depend on both effects: of temperature and temporal evolution (universe expansion).  It is important to note that the structure of the energy-momentum tensor determines coincident terms both in the Stefan-Boltzmann law and in the Casimir effect.  This is because the terms that yield the dependencies of $T^4$ and $d^{- 4}$ are also present in the multiplicative factor of $\xi$.  As future perspectives, we hope to analyze both the Casimir effect and the Stefan-Boltzmann law of a system consisting of a scalar field coupled to a self-contained expanding Universe with periodic spatial conditions.  We suspect that discrete quantities will be obtained for the energy and pressure in both effects.  This could lead to a discrete thermal capacity for the Kasner Universe.

\section*{Acknowledgments}

This work by A. F. S. is supported by CNPq projects 308611/2017-9 and 430194/2018-8.

\end{document}